\documentclass{ws-jhde}

\usepackage{amsmath, amssymb, bbm}

\newcommand{\gfour}{{^{(4)}{\mathbf g}}}
\newcommand{\Lie}{{\cal L}}
\newcommand{\CF}{\psi}
\newcommand{\trK}{K}

\newcommand{\g}{g}
\newcommand{\cg}{\tilde{g}}
\newcommand{\K}{K}
\newcommand{\A}{A}
\newcommand{\cA}{\tilde{A}}
\newcommand{\deriv}{\nabla}
\newcommand{\cderiv}{\tilde{\nabla}}
\newcommand{\Long}{\mathbb{L}}
\newcommand{\cLong}{\tilde{\mathbb{L}}}
\newcommand{\R}{R}
\newcommand{\cR}{\tilde{R}}
\newcommand{\N}{N}
\newcommand{\cN}{\tilde{N}}
\newcommand{\Christoffel}{\Gamma}
\newcommand{\cChristoffel}{\tilde{\Gamma}}
\renewcommand{\u}{u}
\newcommand{\cu}{\tilde{u}}
\newcommand{\s}{\sigma}
\newcommand{\cs}{\tilde{\sigma}}
\newcommand{\cj}{{\tilde{\jmath}\,}}
\newcommand{\crho}{\tilde{\rho}}

\begin{document}
\markboth{Harald P. Pfeiffer}
{The initial value problem in numerical relativity}

%
\catchline{}{}{}{}{}
%

\title{The initial value problem in numerical relativity}

\author{Harald P. Pfeiffer}

\address{Theoretical Astrophysics, California Institute of Technology,
 		 Pasadena, California\ \ 91125}
\maketitle

\begin{history}
\received{(Day Mth. Year)}
\revised{(Day Mth. Year)}
\comby{[editor]}
\end{history}

\begin{abstract}
{\bfseries Abstract.}\quad The conformal method for constructing
initial data for Einstein's equations is presented in both the
Hamiltonian and Lagrangian picture (extrinsic curvature decomposition
and conformal thin sandwich formalism, respectively), and advantages
due to the recent introduction of a weight-function in the extrinsic
curvature decomposition are discussed.  I then describe recent
progress in numerical techniques to solve the resulting elliptic
equations, and explore innovative approaches toward the construction
of astrophysically realistic initial data for binary black hole
simulations.
\end{abstract}

\keywords{Einsteins equations; initial value problem; numerical
relativity.}

\section{Introduction}

Numerical methods play an important role for investigations into the
properties of Einstein's equations.  In particular, the late stages of
inspiral and coalescence of binary compact objects like binary black
holes are thought to be accessible only to numerical investigations.
Knowledge of the full waveform of inspiraling binary black holes,
including the highly nonlinear coalescence phase, will enhance
sensitivity of gravitational wave detectors like LIGO or GEO600
through cross-correlation of the observed signal with the expected
waveforms~\cite{Flanagan-Hughes:1998}.  Comparison of the observed
signals with the predictions of general relativity will test 
general relativity in the genuinely nonlinear regime.  Besides the
experimental urgency, the binary black hole problem is arguably the
most fundamental dynamical problem in general relativity; however, it
remains unsolved.

Initial data forms the starting point for any evolution.  For
Einstein's equations, the most widely used method to construct initial
data is the conformal method, pioneered by
Lichnerowicz~\cite{Lichnerowicz:1944} and extended to a more general
form by York and
coworkers~\cite{York:1972,Murchadha-York:1974b,York:1979}.  In two
recent papers, York~\cite{York:1999} and Pfeiffer \&
York~\cite{Pfeiffer-York:2003} completed the conformal method: It is
now available in a Lagrangian and in a Hamiltonian picture (referred
to as the conformal thin sandwich formalism and the extrinsic curvature
decomposition, respectively), and both pictures completely agree.  The
transverse-tracefree part of the extrinsic curvature is now defined
such that it vanishes for any stationary spacetime. The method is now
completely invariant to conformal transformations of the free data.

The conformal method results in a set of coupled nonlinear
three-dimensional elliptic partial differential equations.  Over the
last few years, numerical techniques for solving these coupled
elliptic equations were improved tremendously.  Construction of binary
black hole initial data is no longer limited by numerical
capabilities, but by the incomplete understanding of the choice of
free data and boundary conditions for the elliptic equations.

Here, we present the Lagrangian and Hamiltonian pictures of the
conformal method, including many details which may have been mentioned
in passing in technical papers, but were never presented in a coherent
fashion.  In the second part of this paper, we describe briefly
numerical methods and then explore recent innovative approaches to the
construction of astrophysically realistic binary black hole initial
data.  Throughout this paper, emphasis is placed on physical and
numerical issues, rather than mathematical proofs.

\section{The initial value problem}
\label{sec:ivp}

Using the standard 3+1 decomposition~\cite{Arnowitt-Deser-Misner:1962,
York:1979} of Einstein's equations, we foliate spacetime with
spacelike $t\!\!=$const.  hypersurfaces.  Each such hypersurface has a
future pointing unit-normal $n^\mu$, induced metric
$g_{\mu\nu}=\gfour_{\mu\nu}+n_\mu n_\nu$ and extrinsic curvature
$K_{\mu\nu} = -\frac{1}{2}\Lie_n g_{\mu\nu}$.  The spacetime metric
can be written as
\begin{equation}\label{eq:spacetime-ds}
ds^2=-\N^2dt+\g_{ij}\left(dx^i+\beta^i dt\right)
\left(dx^j+\beta^jdt\right),
\end{equation}
where $\N$ and $\beta^i$ denote the lapse function and shift vector,
respectively.  $\N$ measures the proper separation between neighboring
hypersurfaces along the surface normals and $\beta^i$ determines how the
coordinate labels move between hypersurfaces: Points along the
integral curves of the time--vector $t^\mu=\N n^\mu + \beta^\mu$
(where $\beta^\mu=[0, \beta^i]$), have the same spatial coordinates
$x^i$.

Einstein's equations decompose into evolution equations and constraint
equations for the quantities $\g_{ij}$ and $\K_{ij}$.  The {\em
evolution equations} determine how $\g_{ij}$ and $\K_{ij}$ are related
between neighboring hypersurfaces,
\begin{align}
\label{eq:dtgij1}
\partial_t\g_{ij}
&=-2\N \K_{ij}+\deriv_i\beta_j+\deriv_j\beta_i\\
\partial_t\K_{ij}
&=\N\left(\R_{ij}-2\K_{ik}\K^k_j+\trK\K_{ij}-8\pi GS_{ij}+4\pi G\g_{ij}(S-\rho)
\right)\nonumber\\
\label{eq:dtKij1}
&\qquad-\deriv_i\deriv_j\N+\beta^k\deriv_k\K_{ij}+\K_{ik}\deriv_j\beta^k
+\K_{kj}\deriv_i\beta^k.
\end{align}
Here, $\deriv_i$ and $\R$ are the covariant derivative and the scalar
curvature (trace of the Ricci tensor) of $\g_{ij}$, respectively, and
$\trK=\K_{ij}\g^{ij}$ denotes the mean curvature.  Furthermore, $G$
stands for Newton's constant, $\rho$ and $S_{ij}$ are matter density and
stress tensor, respectively, and $S=S_{ij}\g^{ij}$ denotes the trace
of $S_{ij}$.

The {\em constraint equations} are conditions within each hypersurface
alone, ensuring that the three-dimensional surface can be embedded
into the four-dimensional spacetime:
\begin{align}
\label{eq:Ham1}
\R+\trK^2-\K_{ij}\K^{ij}&=16\pi G\rho,\\
\label{eq:Mom1}
\deriv_j\left(\K^{ij}-\g^{ij}\trK\right)&=8\pi G j^i,
\end{align}
with $j^i$ denoting the matter momentum density.
Equation~(\ref{eq:Ham1}) is called the {\em Hamiltonian constraint},
and Eq.~(\ref{eq:Mom1}) is the {\em momentum constraint}.

Cauchy initial data for Einstein's equations consists of $(\g_{ij},
\K^{ij})$ on one hypersurface satisfying the constraint
equations~(\ref{eq:Ham1}) and (\ref{eq:Mom1}).  After choosing lapse
and shift (which are arbitrary and merely choose a specific coordinate
system), Eqs.~(\ref{eq:dtgij1}) and (\ref{eq:dtKij1}) determine
$(\g_{ij}, \K^{ij})$ at later times.  Analytically, the constraints
equations are preserved under the evolution.  In practice, however,
during numerical evolution of Eqs.~(\ref{eq:dtgij1}) and
(\ref{eq:dtKij1}) or any other formulation of Einstein's equations,
many problems arise.

The constraints~(\ref{eq:Ham1}) and (\ref{eq:Mom1}) restrict four of
the twelve degrees of freedom of $(\g_{ij}, \K^{ij})$.  As these
equations are not of any standard mathematical form, it is not obvious
which four degrees of freedom are restricted.  Hence, finding any
solutions is not trivial, and it is even harder to construct specific
solutions that represent certain astrophysically relevant situations
like a binary black hole.

\subsection{Preliminaries}
\label{sec:IVP:Preliminaries}

Both Hamiltonian and Lagrangian viewpoints use a conformal
transformation on the spatial metric,
\begin{equation}
\label{eq:g}
\g_{ij}=\CF^4\cg_{ij}
\end{equation}
with strictly positive {\em conformal factor} $\CF$.  $\cg_{ij}$ is
referred to as the {\em conformal metric}.  From (\ref{eq:g}) it
follows that the Christoffel symbols of the physical and conformal
metrics are related by
\begin{equation}\label{eq:Gamma-scaling}
\Christoffel^i_{jk}
=\cChristoffel^i_{jk}+2\CF^{-1}
        \left(\delta^i_j\partial_k\CF+\delta^i_k\partial_j\CF
              -\cg_{jk}\cg^{il}\partial_l\CF\right),
\end{equation}
which in turn implies that the scalar curvatures of $\g_{ij}$ and
$\cg_{ij}$ are related by
\begin{equation}\label{eq:R-scaling}
\R=\CF^{-4}\cR-8\CF^{-5}\cderiv^2\CF.
\end{equation}
Equations (\ref{eq:g})--(\ref{eq:R-scaling}) were already known to
Eisenhart~\cite{Eisenhart:1925}.  Furthermore, for any symmetric
tracefree tensor $\tilde S^{ij}$,
\begin{equation}\label{eq:div-scaling}
\deriv_j\left(\CF^{-10}\tilde S^{ij}\right)=\CF^{-10}\cderiv_j\tilde S^{ij},
\end{equation}
where $\cderiv$ is the covariant derivative of $\cg_{ij}$.
Lichnerowicz~\cite{Lichnerowicz:1944} used Eqs.~(\ref{eq:g}) to
(\ref{eq:div-scaling}) to treat the initial value problem on maximal
slices, $\trK=0$.  For non-maximal slices, the extrinsic curvature is
split into trace and tracefree parts~\cite{Murchadha-York:1974b},
\begin{equation}\label{eq:K-split}
\K^{ij}=\A^{ij}+\frac{1}{3}\g^{ij}\trK.
\end{equation}

With (\ref{eq:R-scaling}) and (\ref{eq:K-split}), the Hamiltonian constraint
(\ref{eq:Ham1}) becomes
\begin{equation}\label{eq:Ham2}
\cderiv^2\CF-\frac{1}{8}\CF\cR-\frac{1}{12}\CF^5\trK^2
+\frac{1}{8}\CF^5\A_{ij}\A^{ij}+2\pi G \CF^{5}\rho=0,
\end{equation}
a quasi-linear Laplace equation for $\CF$.  Local uniqueness proofs of
equations like (\ref{eq:Ham2}) often linearize around an (assumed)
solution, and apply the maximum principle.  However, the signs of the
last two terms of (\ref{eq:Ham2}) are such that the maximum principle
cannot be applied and consequently, it is not immediately guaranteed
that Eq.~(\ref{eq:Ham2}) has locally unique solutions.  The term
proportional to $\A_{ij}\A^{ij}$ will be dealt with later; for the
matter terms we follow York~\cite{York:1979} and introduce conformally
scaled source terms:
\begin{align}
j^i&=\CF^{-10}\cj^i,  \label{eq:j-scaling}\\
\rho&=\CF^{-8}\crho.  \label{eq:rho-scaling}
\end{align}
The scaling for $j^i$ makes the momentum constraint below somewhat
nicer; the scalings of $\rho$ and $j^i$ are tied together such that
the dominant energy condition preserves sign:
\begin{equation}
\rho^2-\g_{ij}j^ij^j
=\CF^{-16}\left(\crho^2-\cg_{ij}\cj^i\cj^j\right)
\ge 0.
\end{equation}
With Eq.~(\ref{eq:rho-scaling}), the matter term in (\ref{eq:Ham2}) becomes
$2\pi G\CF^{-3}\crho$ with negative semi-definite linearization for
$\crho\ge 0$.

The decomposition of $\K_{ij}$ into trace and tracefree part,
Eq.~(\ref{eq:K-split}), turns the momentum constraint~(\ref{eq:Mom1})
into
\begin{equation}\label{eq:Mom2}
\deriv_j\A^{ij}-\frac{2}{3}\deriv^i\trK=8\pi Gj^i.
\end{equation}

The conformal transformation (\ref{eq:g}) implies one additional
conformal scaling relation.  The longitudinal operator
\cite{Deser:1967, York:1973, York:1974}
\begin{equation}\label{eq:L}
(\Long V)^{ij}\equiv\deriv^iV^j+\deriv^jV^i-\frac{2}{3}\g^{ij}\deriv_kV^k,
\end{equation}
satisfies~\cite{York:1973}
\begin{equation}\label{eq:L-scaling}
(\Long V)^{ij}=\CF^{-4}(\cLong V)^{ij}.
\end{equation}
Here $(\cLong V)^{ij}$ is given by the same formula (\ref{eq:L}) but
with quantities associated with the conformal metric $\cg_{ij}$.  (In
fluid dynamics $(\Long V)^{ij}$ is twice the shear of the velocity
field $V^i$).  In $d$ spatial dimensions, the factor $2/3$ in
Eq.~(\ref{eq:L}) is replaced by $2/d$; Eq.~(\ref{eq:L-scaling}) holds
for all $d$.

\subsection{Lagrangian picture --- Conformal thin sandwich formalism}
\label{sec:IVP:CTS}

\begin{figure}
{
\includegraphics{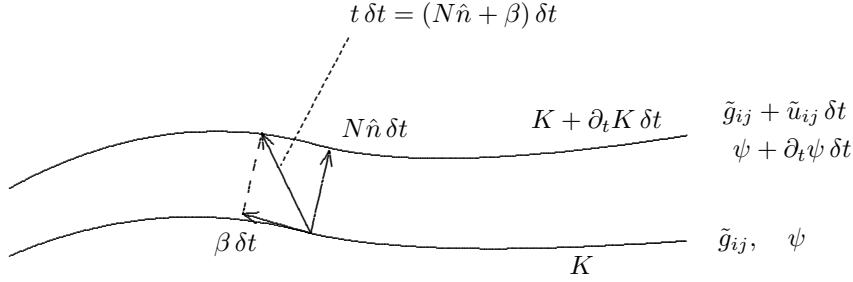}
}
\caption{\label{fig:ConformalThinSandwich}Setup for conformal thin
sandwich formalism.}
\end{figure}

The conformal thin sandwich formalism\cite{York:1999} deals with the
conformal metric and its {\em time derivative}; as illustrated in
Figure~\ref{fig:ConformalThinSandwich}, we deal with {\em two}
hypersurfaces separated by an infinitesimal $\delta t$ (explaining the
name ``thin sandwich''), and connected by lapse $\N$ and shift
$\beta^i$.  The mean curvature of each hypersurface is given by $\trK$
and $\trK+\partial_t\trK\delta t$, respectively, and the metric is
decomposed into conformal factor and conformal metric.  This
decomposition is synchronized between the two hypersurfaces by the
requirement that the conformal metrics on both hypersurfaces have the
same determinant to first order in $\delta t$.  The variation of the
determinant of $\cg_{ij}$ is
\begin{equation}
\delta \cg=\cg\cg^{ij}\delta
\cg_{ij}=\cg\cg^{ij}\cu_{ij}\delta t,
\end{equation}
so that $\cu_{ij}\equiv\partial_t\cg_{ij}$ must be traceless.  

Besides the relationships indicated in Figure
\ref{fig:ConformalThinSandwich}, the conformal thin sandwich formalism
rests on the nontrivial scaling behavior of the lapse function:
\begin{equation}\label{eq:N-scaling}
\N=\CF^6 \cN.
\end{equation}
Indications suggesting this scaling appear in a bewildering variety of
contexts (see discussion in \cite{York:1999, Pfeiffer-York:2003}).
This scaling is also crucial in the present context as well,
cf. Eq.~(\ref{eq:Aij-scaling}) below.

Substitution of Eq.~(\ref{eq:g}) into the evolution equation for
the metric, Eq.~(\ref{eq:dtgij1}), and splitting into trace
and trace-free parts with respect to the physical inverse metric
$\g^{ij}=\CF^{-4}\cg^{ij}$ results in 
\begin{align}
\label{eq:dtphi}
\partial_t\ln\CF&=-\frac{1}{6}\N\trK+\frac{1}{6}\deriv_k\beta^k,\\
\label{eq:uij}
\CF^4\cu_{ij}&=-2\N\A_{ij}+(\Long\beta)_{ij}.
\end{align}
Equation~(\ref{eq:uij}) is the tracefree piece of
$\partial_t\g_{ij}$, thus for $\u_{ij}\equiv \CF^4\cu_{ij}$,
\begin{equation}\label{eq:dtg-tracefree}
\u_{ij}=\partial_t\g_{ij}-\frac{1}{3}\g_{ij}\g^{kl}\partial_t\g_{kl}.
\end{equation}
We solve Eq.~(\ref{eq:uij}) for $\A^{ij}$,
\begin{equation}\label{eq:CTS-Aij}
\A^{ij}=\frac{1}{2\N}\left((\Long\beta)^{ij}-\u^{ij}\right),
\end{equation}
and rewrite with conformal quantities [using (\ref{eq:N-scaling}),
(\ref{eq:L-scaling}) and $\u^{ij}=\CF^{-4}\cu^{ij}$]:
\begin{equation}\label{eq:Aij-scaling}
\A^{ij}
=\CF^{-10}\frac{1}{2\cN}\Big((\cLong\beta)^{ij}-\cu^{ij}\Big)
\equiv\CF^{-10}\cA^{ij},
\end{equation}
which defines the conformal tracefree extrinsic curvature $\cA^{ij}$.
Equation~(\ref{eq:Aij-scaling}) shows that the formula for $\A^{ij}$
is form invariant under conformal transformations; this hinges on the
scaling of $\N$ in Eq.~(\ref{eq:N-scaling}).  Substitution of
Eq.~(\ref{eq:Aij-scaling}) into the momentum constraint
(\ref{eq:Mom2}) and application of Eq.~(\ref{eq:div-scaling}) yields
\begin{equation}\label{eq:Mom3}
\cderiv_j\left(\frac{1}{2\cN}(\cLong\beta)^{ij}\right)
-\cderiv_j\left(\frac{1}{2\cN}\cu^{ij}\right)-\frac{2}{3}\CF^6\cderiv^i\trK
=8\pi G\cj^i,
\end{equation}
whereas Eq.~(\ref{eq:Aij-scaling}) modifies the Hamiltonian constraint
(\ref{eq:Ham2}) to
\begin{equation}\label{eq:Ham3}
\cderiv^2\CF-\frac{1}{8}\CF \cR-\frac{1}{12}\CF^5\trK^2
+\frac{1}{8}\CF^{-7}\cA_{ij}\cA^{ij}=-2\pi G\CF^{-3}\crho.
\end{equation}
Equations~(\ref{eq:Mom3}) and~(\ref{eq:Ham3}) constitute elliptic
equations for $\beta^i$ and $\CF$.  We can therefore construct a valid
initial data set as follows: Choose the free data
\begin{equation}\label{eq:CTS-freedata1}
(\cg_{ij},\; \cu_{ij},\;\;\;\trK,\; \cN)
\end{equation}
(and matter terms if applicable), solve Eqs.~(\ref{eq:Mom3}) and
(\ref{eq:Ham3}) for $\beta^i$ and $\CF$, and finally, assemble
$\g_{ij}=\CF^4\cg_{ij}$ and
$\K^{ij}=\CF^{-10}\cA^{ij}+\frac{1}{3}\g^{ij}\trK$. 

We now comment on several issues related to the conformal thin
sandwich formalism.

\subsubsection{Fixing $\cN$ via $\partial_t\trK$}

In the free data Eq.~(\ref{eq:CTS-freedata1}), $\cg_{ij}$ and
$\cu_{ij}=\partial_t\cg_{ij}$ constitute a ``variable~\&~velocity
pair'' $(q, \dot q)$ in the spirit of Lagrangian mechanics, but
the remaining free data does not.  To improve this situation, we note
that the trace of the evolution equation~(\ref{eq:dtKij1}), results in
(see, e.g., \cite{Smarr-York:1978b})
\begin{equation}\label{eq:dtK}
\partial_tK-\beta^k\partial_k\trK
=\N\left(\R+\trK^2+4\pi G(S-3\rho)\right)-\deriv^2\N.
\end{equation}
Elimination of $\R$ with the Hamiltonian constraint (\ref{eq:Ham1})
and rewriting this equation with conformal quantities results in
\begin{align}\label{eq:dtK3}
\cderiv^2\cN+14\cderiv^i&\ln\CF\cderiv_{\!i}\cN
+\cN\bigg[\frac{3}{4}\cR+\frac{1}{6}\CF^4\trK^2
          \!-\!\frac{7}{4}\CF^{-8}\cA_{ij}\cA^{ij}\\
+&42\cderiv_{\!i}\ln\CF\cderiv^i\ln\CF\!-\!4\pi G\CF^4(S\!+\!4\rho)\bigg]
=-\CF^{-2}\left(\partial_t\trK\!-\!\beta^k\partial_k\trK\right).
\nonumber
\end{align}
For given $\partial_tK$, this constitutes an elliptic equation for
$\tilde N$.  Therefore, if we take $\partial_t\trK$ as the ``free'' quantity
instead of $\cN$, then the free data for the conformal thin sandwich
formalism becomes
\begin{equation}\label{eq:CTS-freedata2}
(\cg_{ij},\; \cu_{ij},\;\;\;\trK,\;\partial_t\trK)
\end{equation}
plus matter terms if applicable.  These free data consist completely
of $(q, \dot q)$ pairs as appropriate for the Lagrangian viewpoint.
These free data are also useful in practice for computations of
quasi-equilibrium initial data, for which $\partial_t\trK=0$ is a
natural and simple choice, whereas it is not obvious which conformal
lapse $\cN$ should be used.  We note that Eq.~(\ref{eq:dtK3}) can be
rewritten as
\begin{align}\label{eq:dtK4}
\cderiv^2(\cN\CF^7)-(\cN\CF^7)\bigg[\frac{1}{8}\cR+\frac{5}{12}\CF^4\trK^2
+\frac{7}{8}\CF^{-8}\cA_{ij}\cA^{ij}&+2\pi G\CF^4(\rho+2S)\bigg]\\
&=-\CF^5\left(\partial_t\trK\!-\!\beta^k\partial_k\trK\right).
\nonumber
\end{align}
This equation is somewhat shorter and computationally somewhat more
convenient.  Using the free data (\ref{eq:CTS-freedata2}), {\em five}
coupled elliptic equations have to be solved, rather than four, namely
(\ref{eq:Mom3}), (\ref{eq:Ham3}) and either (\ref{eq:dtK3}) or
(\ref{eq:dtK4}).

\subsubsection{Invariance to conformal transformations of the free data}
\label{sec:IVP:CTS:invariance}

Given free data (\ref{eq:CTS-freedata1}) and a solution $(\CF,
\beta^i)$ of the conformal thin sandwich equations, choose a function
$\Psi>0$, and define conformally rescaled free data by
\begin{equation}\label{eq:freedata-scaled}
\cg'_{ij}=\Psi^{-4}\cg_{ij},\quad
{\cu{'}{\,}}^{ij}=\Psi^{4}\cu^{ij},\quad
K'=K,\quad
\cN'=\Psi^{-6}\cN,
\end{equation}
plus the scalings $\crho'=\Psi^8\crho,\quad
{\cj'}{\,}^i=\Psi^{10}\cj^i$ for matter terms if applicable.  These
free data, together with conformal factor $\CF'=\Psi\CF$ and the shift
$\beta'{\,}^i=\beta^i$ 
lead to the {\em same} physical initial data $(\g_{ij}, \K^{ij})$:
\begin{align}
\g'_{ij}
&=\CF'{\,}^4\;\cg'_{ij}=\CF^4\cg_{ij}=\g_{ij},\\
\A'{\,}^{ij}
&=\CF'{\,}^{-10}\frac{1}{2\cN'}
\left(\big(\cLong'\,\beta'\big)^{ij}-\cu'{\,}^{ij}\right)
\label{eq:Aij-conformal-invariance}
=\A^{ij}.
\end{align}
Here, we used Eqs.~(\ref{eq:g}) and~(\ref{eq:Aij-scaling}), and
$\cLong'$ denotes the longitudinal operator of $\cg'_{ij}$, which, by
Eq.~(\ref{eq:L-scaling}), satisfies
$(\cLong'\beta)^{ij}=\Psi^4(\cLong\beta)^{ij}$.  Adding the trace of
the extrinsic curvature to Eq.~(\ref{eq:Aij-conformal-invariance}) is
trivial.

Therefore, only the {\em conformal equivalence class} of $\cg_{ij}$ is
relevant for the physical solution.  This is a very desirable
property; we introduced $\cg_{ij}$ as a {\em conformal} metric, so its
overall scaling should not matter.  Specification of $\partial_t\trK$
instead of $\cN$ as part of the free data preserves this invariance,
as the $\partial_t\trK$-equation, (\ref{eq:dtK3}), is derived from
{\em physical} quantities in the first place.

The extrinsic curvature decomposition introduced in the next section
is also invariant under conformal transformations of the free data.
We note that invariance under conformal transformations of the free
data is not trivial; earlier variants of the constraint decompositions
did not possess it, giving rise to ambiguities in the free data which
influenced numerical investigations~\cite{Tichy-Bruegmann-etal:2003}.

\subsubsection{Gauge degrees of freedom}

The physical initial data $(\g_{ij}, \K^{ij})$ has twelve degrees of
freedom (matter just adds four additional degrees of freedom in
$\crho$ and $\cj^i$ which determine the four physical matter variables
$\rho$ and $j^i$).  It is constrained by four constraint equations, so
there should be eight degrees of freedom in the freely specifiable
data.  However, even taking into account that only the conformal
equivalence class of $\cg_{ij}$ is relevant and that $\cu_{ij}$ is
traceless, the free data (\ref{eq:CTS-freedata1}) or
(\ref{eq:CTS-freedata2}) consists of {\em twelve} quantities, not
eight.  To clarify this issue, consider the substitutions
\begin{equation}\label{eq:beta-gauge}
\begin{aligned}
\cu^{ij}\;&\to\;\cu^{ij}+(\cLong W)^{ij},\\
\beta^i\;&\to\;\beta^i+W^i.
\end{aligned}
\end{equation}
The vector $W^i$ disappears from
Eqs.~(\ref{eq:CTS-Aij})--(\ref{eq:Ham3}), therefore the substitution
(\ref{eq:beta-gauge}) will not change the physical initial data set
$(\g_{ij}, \K^{ij})$; it merely tilts the time-axis and changes the
coordinate labels on the second hypersurface.  Thus, $\cu_{ij}$
contains three gauge degrees of freedom associated with the shift.
($W^i$ enters into Eq.~(\ref{eq:dtK3}) as an advection term, though,
because $\partial_tK$ is the derivative along the time-vector).

The fourth ``missing'' degree of freedom is hidden in the lapse
function $\cN$:  One can construct every possible initial data set
$(\g_{ij}, \K^{ij})$ with {\em any} (non-pathologic) choice of $\cN$.
This can be seen by going backward from the physical initial data
$(\g_{ij}, \K^{ij})$ (satisfying the constraints) to the free data.
Given $(\g_{ij}, \K^{ij})$ {\em and} any $\cN$ and $\beta^i$,
set the free data (\ref{eq:CTS-freedata1}) by
\begin{equation}
\cg_{ij}=\g_{ij},\quad \cu^{ij}=(\Long\beta)^{ij}-2\cN \A^{ij},\quad 
\trK=\K^{ij}\g_{ij}
\end{equation}
as well as the given $\cN$.  With these free data, $\CF\equiv 1$ and
the given $\beta^i$ will reconstruct the physical spacetime $(\g_{ij},
\K^{ij})$ as can be seen from Eqs.~(\ref{eq:g}) and
(\ref{eq:Aij-scaling}).  Therefore, $\CF\equiv 1$ and the given
$\beta^i$ will solve the conformal thin sandwich equations
(\ref{eq:Mom3}) and (\ref{eq:Ham3}).

The fact that we were free to choose $\beta^i$ reflects again the
gauge-symmetry illustrated in Eq.~(\ref{eq:beta-gauge}), but in
addition, we showed that the choice of $\cN$ does not restrict the set
of ``reachable'' initial data sets.

Note that the physical initial data contain further gauge freedom:
Covariance under spatial transformations implies that $\g_{ij}$ (and
$\cg_{ij}$) contain three gauge degrees of freedom associated with the
choice of coordinates.  Furthermore, $\trK$ can be interpreted as {\em
time}~\cite{York:1972}, fixing the temporal gauge.  Thus, an initial
data set has only four physical degrees of freedom---in perturbed flat
space they are simply the two polarizations of gravitational waves.

\subsubsection{Implications for an evolution of the initial data}
\label{sec:IVP:ImplicationsForEvolution}

During the solution of the conformal thin sandwich equations, one
finds a shift $\beta^i$ and a lapse $N$.  If this gauge is used in a
subsequent evolution of the initial data $(\g_{ij}, \K^{ij})$ then
Eq.~(\ref{eq:dtg-tracefree}) implies that, {\em initially},
\begin{equation}\label{eq:dtg-tracefree2}
\partial_t\g_{ij}-\frac{1}{3}\g_{ij}\g^{kl}\partial_t\g_{kl}
=\CF^4\cu_{ij}.
\end{equation}
The freely specifiable piece $\cu_{ij}$ thus directly controls the
tracefree part of the time-derivative of the metric.  If we specified
$\partial_t\trK$ as part of the free data, then, of course, this will
be the initial time-derivative of the mean curvature.  Finally, from
(\ref{eq:dtphi}), we find
\begin{align}
\partial_t\ln\CF&=\frac{1}{6}\left(\deriv_k\beta^k-\N\trK\right)\nonumber\\
\label{eq:dtphi2}
&=\frac{1}{6}\cderiv_k\beta^k+\beta^k\partial_k\ln\CF-\frac{1}{6}\CF^6\cN\trK.
\end{align}

Since we have been very successful so far with specification of
time-derivatives ($\partial_t\cg_{ij}$ and $\partial_t\trK$), one
might be tempted to turn (\ref{eq:dtphi2}) around and use it as the
{\em definition} of $\trK$ in terms of $\partial_t\ln\CF$.  This idea
certainly comes to mind when looking for quasi-equilibrium solutions,
for which time-derivatives of as many quantities as possible should
vanish.  Pursuing this idea, we find from Eq.~(\ref{eq:dtphi2})
\begin{equation}\label{eq:K-attempt}
K=\frac{1}{\CF^6\cN}\left(\cderiv_k\beta^k
       -6\left(\partial_t-\beta^k\partial_k\right)\ln\CF\right).
\end{equation}
Substituting Eq.~(\ref{eq:K-attempt}) into the momentum constraint
Eq.~(\ref{eq:Mom3}), however, makes the principal part of
Eq.~(\ref{eq:Mom3}) proportional to
\begin{equation}
\partial_j\partial^j\beta^i-\partial^i\partial_k\beta^k,
\end{equation}
which is non-invertible.  Therefore the attempt to fix $\trK$ via
(\ref{eq:K-attempt}) will fail.  Indeed, during the construction
quasi-equilibrium initial data sets of spherically symmetric
spacetimes~\cite{Cook-Pfeiffer:2004}, it was found that the conformal
thin sandwich formalism with $\cu_{ij}=0$ and $\partial_t\trK=0$ (and
appropriate boundary conditions) is so successful in picking out the
time-like Killing vector that several different choices of $\trK$ lead
to solutions satisfying $\partial_t\ln\CF=0$.  

We conclude that, in contrast to the trace-free part
(\ref{eq:dtg-tracefree2}), one can {\em not} easily control
$\partial_t\ln\CF$ by choices of the free data.  One can only evaluate
(\ref{eq:dtphi2}) {\em after} solving the conformal thin sandwich
equations.

\subsection{Hamiltonian picture --- Extrinsic curvature decomposition}
\label{sec:IVP:ExCurv}

The second method to construct solutions of the constraint equations
if based on a decomposition of the extrinsic curvature.  Early
variants of this approach \cite{Murchadha-York:1974b, York:1979} have
been widely used for almost thirty years, but the final version was
developed only very recently \cite{Pfeiffer-York:2003}.  We will make
use of the equations and results from section
\ref{sec:IVP:Preliminaries}, in particular, we use a conformal metric,
$\g_{ij}=\CF^4\cg_{ij}$, and split the extrinsic curvature into trace
and trace-free parts, $\K^{ij}=\A^{ij}+1/3\g^{ij}\trK$,
cf. Eqs.~(\ref{eq:g}) and (\ref{eq:K-split}).

We start with a {\em weighted transverse traceless} decomposition of
$\A^{ij}$,
\begin{equation}\label{eq:barAij-TT}
\A^{ij}=\A^{ij}_{TT}+\frac{1}{\s}\left(\Long V\right)^{ij}.
\end{equation}
Here, $\A^{ij}_{TT}$ is transverse, $\deriv_{\!j}\A^{ij}_{TT}=0$, and
traceless, $\g_{ij}\A^{ij}_{TT}=0$, and $\s$ is a strictly positive
and bounded function.  Appearance of the weight function $\s$ is a key
point in the extrinsic curvature formulation; its inclusion is the
major difference of~\cite{Pfeiffer-York:2003} over the older variants.

Given a symmetric tracefree tensor like $\A^{ij}$, the decomposition
(\ref{eq:barAij-TT}) is obtained by taking the divergence of
Eq.~(\ref{eq:barAij-TT}),
\begin{equation}\label{eq:div-Aij-TT}
\deriv_j\A^{ij}=\deriv_j\left[\s^{-1}(\Long V)^{ij}\right].
\end{equation}
The right hand side, $\deriv_j\left[\s^{-1}(\Long\,.\,)^{ij}\right]$,
is a well-behaved elliptic operator in divergence form, so no problem
should arise when solving (\ref{eq:div-Aij-TT}) for $V^i$.
Substitution of the solution $V^i$ back into (\ref{eq:barAij-TT})
yields $\A^{ij}_{TT}$.  In the presence of boundaries,
Eq.~(\ref{eq:div-Aij-TT}) requires boundary conditions which will
influence the solution $V^i$ and the decomposition
(\ref{eq:barAij-TT}).  For closed manifolds, existence and uniqueness
of the decomposition (\ref{eq:barAij-TT}) for the case $\s\equiv 1$
was shown in \cite{York:1974}.

We now conformally scale the quantities on the right hand side of
(\ref{eq:barAij-TT}) with the goal of rewriting the momentum
constraint in conformal space.  First, we set
\begin{equation}\label{eq:AijTT-scaling}
\A^{ij}_{TT}\equiv \CF^{-10}\cA^{ij}_{TT}.
\end{equation}
Equation~(\ref{eq:div-scaling}) ensures that $\cA^{ij}_{TT}$ is
transverse with respect to $\cg_{ij}$ if and only if $\A^{ij}_{TT}$ is
transverse with respect to the physical metric $\g_{ij}$.
Because of Eq.~(\ref{eq:L-scaling}), and because $\Long$ is the
conformal Killing operator, the vector $V^i$ is not rescaled.
The conformal scaling of the weight function is given by
\begin{equation}\label{eq:sigma-scaling}
\s=\CF^6\cs.
\end{equation}
The most immediate reason for this scaling is to allow
Eq.~(\ref{eq:Aij-scaling2}) below; several more reasons will be
mentioned in the sequel.

Using the scaling relations~(\ref{eq:L-scaling}),
 (\ref{eq:AijTT-scaling}) and~(\ref{eq:sigma-scaling}), we can rewrite
 Eq.~(\ref{eq:barAij-TT}) as
\begin{equation}\label{eq:Aij-scaling2}
\A^{ij}
=\CF^{-10}\left(\cA^{ij}_{TT}+\frac{1}{\cs}(\cLong V)^{ij}\right)
=\CF^{-10}\cA^{ij},
\end{equation}
where
\begin{equation}\label{eq:Aij-TT}
\cA^{ij}\equiv\cA^{ij}_{TT}+\cs^{-1}(\cLong V)^{ij}
\end{equation}
is a weighted transverse traceless decomposition in the conformal
space.  The scaling $\A^{ij}=\CF^{-10}\cA^{ij}$ was not postulated (as
it had to be in the old variants), but follows from the other
scalings.  By virtue of the scaling of the weight function $\sigma$,
Eq.~(\ref{eq:sigma-scaling}), the weighted transverse traceless
decomposition thus commutes with the conformal transformation.  This
commutation of conformal transformation and weighted transverse
traceless decomposition is precisely the new feature of the weighted
decomposition.  Without the weight-function, conformal transformations
and transverse-traceless decomposition do not commute, leading to the
two inequivalent old variants, depending on which operation is
performed first.

Equations~(\ref{eq:div-scaling}) and (\ref{eq:Aij-scaling2}) allow us to 
rewrite the momentum constraint (\ref{eq:Mom2}) as
\begin{equation}\label{eq:Mom4}
\cderiv_j\left(\frac{1}{\cs}(\cLong V)^{ij}\right)
-\frac{2}{3}\CF^6\cderiv^i\trK
=8\pi G\cj^i,
\end{equation}
an elliptic equation for $V^i$.  The Hamiltonian constraint
Eq.~(\ref{eq:Ham2}) reads
\begin{equation}\label{eq:Ham4}
\cderiv^2\CF-\frac{1}{8}\cR\CF
-\frac{1}{12}\CF^5\trK^2+\frac{1}{8}\CF^{-7}\cA_{ij}\cA^{ij}
=-2\pi G\CF^{-3}\crho,
\end{equation}
with $\cA^{ij}$ given by Eq.~(\ref{eq:Aij-TT}).  Equation~(\ref{eq:Ham4})
is identical to Eq.~(\ref{eq:Ham3}) since in both formulations
$\A^{ij}=\CF^{-10}\cA^{ij}$, however, the definitions of $\cA^{ij}$
differ.

Starting from the physical initial data $(\g_{ij}, \K^{ij})$, we have
now rewritten the constraints (\ref{eq:Ham1}) and (\ref{eq:Mom1}) as
elliptic equations (\ref{eq:Mom4}) and (\ref{eq:Ham4}).  In order to
construct a valid initial data set $(\g_{ij}, \K^{ij})$, one first
chooses the free data
\begin{equation}\label{eq:ExCurv-freedata}
\left(\cg_{ij},\;\; \trK,\;\; \cA^{ij}_{TT},\;\;\cs\right)
\end{equation}
and matter terms if applicable, then solves Eqs.~(\ref{eq:Mom4}) and
(\ref{eq:Ham4}) for $V^i$ and $\CF$, and finally assembles the
physical solution by Eqs.~(\ref{eq:g}), (\ref{eq:K-split}), and
(\ref{eq:Aij-scaling2}).

\subsubsection{Remarks on the extrinsic curvature decomposition}

Similar to section \ref{sec:IVP:CTS:invariance}, one can show that the
physical initial data $(\g_{ij}, \K^{ij})$ is {\em invariant} to 
a conformal transformation of the free data.  For $\Psi>0$, the relevant
transformations are [cf. Eq.~(\ref{eq:freedata-scaled})]:
\begin{equation}\label{eq:freedata-scaled2}
\cg'_{ij}=\Psi^{-4}\cg_{ij},\quad
\cA'{\,}^{ij}_{TT}=\Psi^{10}A^{ij}_{TT},\quad
K'=K,\quad
\cs'=\Psi^{-6}\cs,
\end{equation}
plus the scalings $\crho'\!=\!\Psi^8\crho,\quad
{\cj'}{\,}^i\!=\!\Psi^{10}\cj^i$ for matter terms if applicable.  The
calculation is straightforward, the key-point being that the scaling of
the weight-function~(\ref{eq:sigma-scaling}) synchronizes the
conformal scaling of the transverse-traceless and longitudinal parts
of the weighted transverse traceless decomposition.

Because of the invariance to conformal scalings of the free data,
$\cg_{ij}$ supplies only five degrees of freedom, so that the free
data Eq.~(\ref{eq:ExCurv-freedata}) contains nine degrees of freedom.
The weight $\s$ (or $\cs$) merely parametrizes the transverse
traceless decomposition (\ref{eq:barAij-TT}).  For any choice of $\s$,
the decomposition (\ref{eq:barAij-TT}) can be performed, therefore
with any choice of $\cs$, all initial data sets can be generated for
appropriate choices of the free data.

To construct a transverse traceless tensor $\cA^{ij}_{TT}$ compatible
with the metric $\cg_{ij}$, one decomposes a general symmetric
tracefree tensor $\tilde M^{ij}$.  Write
\begin{equation}\label{eq:Mij-decomposition}
\tilde M^{ij}=\cA^{ij}_{TT}+\frac{1}{\cs}(\cLong W)^{ij}.
\end{equation}
The divergence of this equation, 
\begin{equation}\label{eq:divergence-Mij-decomposition}
\cderiv_j\tilde M^{ij}=\cderiv_j\left[\cs^{-1}(\cLong W)^{ij}\right],
\end{equation}
represents an elliptic equation for $W^i$.  Solving this equation, and
substituting $W^i$ back into (\ref{eq:Mij-decomposition}) yields
\begin{equation}
\cA^{ij}_{TT}=\tilde M^{ij}-\frac{1}{\cs}(\cLong W)^{ij}.
\end{equation}
The formula for $\cA^{ij}$, Eq.~(\ref{eq:Aij-TT}), now reads
\begin{equation}\label{eq:Aij-from-Mij}
\cA^{ij}=\tilde M^{ij}+\frac{1}{\cs}\big[\cLong(V-W)\big]^{ij},
\end{equation}
which depends only on the {\em difference} $V^i-W^i$.  On the other hand, 
subtraction of (\ref{eq:divergence-Mij-decomposition}) from 
the momentum constraint (\ref{eq:Mom4}) yields
\begin{equation}\label{eq:Mom5}
\cderiv_j\left(\frac{1}{\cs}\big[\cLong(V-W)\big]^{ij}\right)
+\cderiv_j\tilde M^{ij}-\frac{2}{3}\CF^6\cderiv^i\trK
=8\pi G\cj^i,
\end{equation}
which is an equation for the {\em difference} $V^i-W^i$.  Thus one can
combine the construction of $\cA^{ij}_{TT}$ from $\tilde M^{ij}$ with
the solution of the momentum constraint, as observed by
Cantor~\cite{York:Cantor-info}.  Instead of solving
(\ref{eq:divergence-Mij-decomposition}) for $W^i$ and then
(\ref{eq:Mom4}) for $V^i$, one can directly solve (\ref{eq:Mom5}) for
$V^i-W^i$.

In the presence of boundaries, solutions of elliptic equations like
(\ref{eq:Mom5}) or (\ref{eq:divergence-Mij-decomposition}) will depend
on {\em boundary conditions}.  When constructing black hole initial
data, inner boundaries are often present, and it is far from clear
what boundary conditions to apply there.  In
Ref.~\cite{Pfeiffer-Cook-Teukolsky:2002}, for example, the situation
is encountered that boundary conditions for the combined solution for
$V^i-W^i$ are known, but not for the individual solutions for $W^i$
and $V^i$.

\subsubsection{Identification of $\s$ with the lapse $\N$}

The extrinsic curvature formulation of the initial value problem as
presented so far is perfectly adequate for the mathematical task of
rewriting the constraints as well-defined equations.  However, it is
very natural to further identify the weight-function $\s$ with the
lapse function $\N$,
\begin{equation}\label{eq:sigma=2N}
\s=2\N,\qquad\cs=2\cN.
\end{equation}
One reason for this identification is that $\s$ and $\N$ have the same
conformal scaling behavior, cf. Eqs.~(\ref{eq:N-scaling}) and
(\ref{eq:sigma-scaling}).  A second reason is that with this
identification, the conformal thin sandwich equations become
equivalent to the extrinsic curvature formulation.  To see this,
note that by virtue of (\ref{eq:sigma=2N}), Eqs.~(\ref{eq:Aij-from-Mij}) 
and (\ref{eq:Mom5}) become
\begin{gather}\label{eq:cA5}
\cA^{ij}=\tilde M^{ij}+\frac{1}{2\cN}\left[\cLong(V-W)\right]^{ij},
\intertext{and}
\label{eq:Mom6}
\cderiv_j\left(\frac{1}{2\cN}\big[\cLong(V-W)\big]^{ij}\right)
+\cderiv_j\tilde M^{ij}-\frac{2}{3}\CF^6\cderiv^i\trK
=8\pi G\cj^i.
\end{gather}
With the identifications
\begin{equation}
\tilde M^{ij}\;\leftrightarrow\;-\frac{1}{2\cN}\cu^{ij},\qquad\quad
V^i-W^i\;\leftrightarrow\; \beta^i,
\end{equation}
Eqs.~(\ref{eq:cA5}) and (\ref{eq:Mom6}) are {\em identical} to
Eqs.~(\ref{eq:Aij-scaling}) and (\ref{eq:Mom3}) of the conformal thin
sandwich formalism.  The Lagrangian picture agrees completely with the
Hamilton\-ian picture.  A third reason for (\ref{eq:sigma=2N}) is
given next.

\subsubsection{Stationary spacetimes have $\A^{ij}_{TT}=0$}

Consider a stationary solution of Einstein's equations with timelike
Killing vector $l$.  Given a spacelike hypersurface $\Sigma$, there is
a preferred gauge so that the time-vector of an evolution coincides
with $l$, namely $\N=-n\cdot l$, $\beta=\,\perp\! l$, where $n$ is the
unit normal to $\Sigma$, and $\perp$ is the projection operator into
$\Sigma$.  With this choice of lapse and shift, $\g_{ij}$ and
$\K^{ij}$ will be time-independent.  Using $\partial_t\g_{ij}=0$ in
Eq.~(\ref{eq:dtgij1}) and taking the tracefree part yields
\begin{equation}
  \A^{ij}=\frac{1}{2\N}(\Long\beta)^{ij},
\end{equation}
a weighted transverse traceless decomposition with $\A^{ij}_{TT}\equiv
0$.  Thus, with the appropriate weight factor $\s=2\N$, the extrinsic
curvature has {\em no transverse traceless piece} for any spacelike
slice in any spacetime with timelike Killing vector (A similar
argument is applicable in the ergosphere of a Kerr black hole;
however, one must be more careful with the choice of $\Sigma$ relative
to $l$).

This is an important result.  One generally identifies the transverse
traceless piece of the extrinsic curvature with {\em radiative degrees
of freedom}.  Stationary spacetimes do not radiate, and therefore
$\A^{ij}_{TT}$ should indeed vanish.  In contrast, a
transverse-traceless decomposition of $\A^{ij}$ without the
weight-factor will in general lead to a nonzero transverse traceless
piece so that such a decomposition is incompatible with the
identification of $\A^{ij}_{TT}$ with ``gravitational radiation.''

\section{Binary black hole initial data}

Up to about five years ago, many assumptions were necessary to
simplify the initial value problem sufficiently to make it tractable
for the computational methods of that time.  
The major assumptions were (see \cite{Cook:2000} for a review)
\begin{enumerate}
\item Maximal slicing, $\trK=0$.
\item Conformal flatness $\cg_{ij}=f_{ij}$, where $f_{ij}$ represents
the Euclidean metric.
\item Use of the analytical Bowen-York~\cite{Bowen-York:1980}
extrinsic curvature to solve the momentum constraint.
\end{enumerate}
Under these assumptions, only a single quasi-linear (flat-space)
Laplace-equation must to be solved numerically for the conformal
factor.  This was done, e.g., with inversion symmetry boundary
conditions~\cite{Cook:1991, Cook-Choptuik-etal:1993} or with the
puncture method~\cite{Brandt-Bruegmann:1997}.  Since then, the
conformal method (as presented in Sec.~\ref{sec:ivp}) was completed,
and several numerical codes were developed that are capable of solving
the coupled constraint equations~\cite{Marronetti-Matzner:2000,
Grandclement-Gourgoulhon-Bonazzola:2001b, Holst:2001,
Pfeiffer-Kidder-etal:2003, Tichy-Bruegmann-etal:2003,
Yo-Cook-etal:grqc0406020}.  Three different numerical techniques have
been used to discretize the elliptic problems, finite differences,
spectral methods and finite elements.  

{\em Finite differences}~\cite{Marronetti-Matzner:2000,
Tichy-Bruegmann-etal:2003, Yo-Cook-etal:grqc0406020} are familiar to
almost all researchers, and are fairly straightforward to implement.
However, the presence of different length scales
\begin{equation}\label{eq:length-scales}
M_{A,B} \ll d \ll R,
\end{equation}
where $M_{A,B}$ represents the masses of the two black holes (labeled
$A$ and $B$), $d$ their separation and $R$ the distance to the outer
boundary of the computational grid, restricts finite difference codes
on uniform grids to a very coarse resolutions with errors of $10^{-2}$
to $10^{-3}$.  Imposing boundary conditions on spherical inner
boundaries is also difficult in finite difference codes.  Adaptive
mesh-refinement may drastically improve the ability of
finite-difference codes to handle the different length scales in
Eq.~(\ref{eq:length-scales}). In Refs.~\cite{Diener-Jansen-etal:2000,
Pretorius:grqc0407110,Brown-Lowe:grqc0411112} the Hamiltonian
constraint alone is solved, and work is in progress to extend adaptive
mesh refinement to the coupled initial value
equations~\cite{Pretorius:2004}.

{\em Spectral methods}~\cite{Grandclement-Gourgoulhon-Bonazzola:2001b,
Pfeiffer-Kidder-etal:2003} (see \cite{Canuto-Hussaini,Boyd:2001} for
general introductions) approximate the solution with a truncated
expansion in some basis functions, typically Chebyshev-polynomials or
spherical harmonics.  The solutions to the constraint equations are
smooth, so that the accuracy improves exponential with the number of
basis functions and much higher accuracies are achieved ($\sim
10^{-9}$ for binary black hole initial data
in~\cite{Pfeiffer-Kidder-etal:2003}).  Spectral methods are also more
efficient, therefore permitting much larger parameter studies, allow
more easily to impose boundary conditions on spherical boundaries (for
expansions in spherical harmonics), and, in combination with
domain-decomposition techniques, easily resolve the different length
scales in Eq.~(\ref{eq:length-scales}).  The most accurate (for binary
black holes) and versatile code seems to be the one developed by the
author~\cite{Pfeiffer-Kidder-etal:2003}, which has been used to solve
essentially all versions of the initial value
problem~\cite{Pfeiffer-Cook-Teukolsky:2002, Pfeiffer:2003,
Cook-Pfeiffer:2004,Pfeiffer-Kidder-etal:grqc0410016}.

{\em Finite elements}~\cite{Holst:2001,Bank-Holst:2003} cover the
computational domain with very many small computational cells
(typically tedrahedra), and expand the solution to low polynomial
order (often just linear) in each cell.  The method presented in
Refs.~\cite{Holst:2001,Bank-Holst:2003} is capable of solving the
coupled constraint equations, and work toward using it for physically
meaningful initial data is under way.

With these new codes and with the final conformal method it is
possible to move beyond the assumptions mentioned above.  In
particular, three separate issues have been pursued: Exploration of
{\em conformally non-flat three-geometries}, which is motivated by the
facts that the Kerr spacetime does not admit conformally flat
slices~\cite{Monroe:1976,Garat-Price:2000, Kroon:2004}, and that a
binary compact object is not conformally flat at second post-Newtonian
order~\cite{Rieth:1997}.  {\em Replacement of the Bowen-York extrinsic
curvature}, which does not exactly reproduce a stationary spinning or
boosted black hole, and which may be responsible for unexpected
behavior of sequences of circular orbits for spinning equal-mass black
holes~\cite{Pfeiffer-Teukolsky-Cook:2000} or irrotational black holes
in the test-mass limit~\cite{Pfeiffer:2003}.  And finally, {\em
physically motivated} boundary conditions at inner excision surfaces
surrounding the singularities of the black holes.

The goal of these new investigations is the construction of
astrophysically realistic binary black hole data, i.e. configurations
as they occur during the inspiral of two black holes in nature.
Ideally, of course, such an initial data set should contain the
outgoing gravitational wave signal of the preceding inspiral.
However, at the current stage of sophistication, this is not taken
into account, and the intermediate goal is to construct initial data
with as little ``spurious,'' ``unphysical'' gravitational energy
content as possible.  The quotes used around the terms ``spurious''
and ``unphysical'' indicate part of the challenge: It is not even
clear what these terms mean precisely on a single spacelike surface.

Some of the research discussed in this section predates the conformal
thin sandwich formalism, or the final extrinsic curvature
decomposition as discussed in Sec.~\ref{sec:ivp}.  These earlier
papers typically use special cases of these more general frameworks,
and we will discuss them from that perspective.

\subsection{Numerical solution of the extrinsic curvature decomposition}

As mentioned above, the Kerr spacetime does not admit conformally flat
slices, so that any initial data built on conformal flatness will not
be able to reproduce Kerr exactly.  One approach to address this issue
---proposed by Matzner {\em et
al.}~\cite{Matzner-Huq-Shoemaker:1999}--- superposes exact quantities
for single (spinning or boosted) black holes to define the free data
for the extrinsic curvature decomposition.  Because the method is
built on analytical single black hole solutions, it trivially works
for single black holes.  Specifically, Matzner {\em et al.}
considered the Kerr-Schild form of the Kerr spacetime,
\begin{equation}\label{eq:KS}
\gfour_{\mu\nu}=\eta_{\mu\nu}+2H l_\mu l_\nu,
\end{equation}
where $\eta_{\mu\nu}$ represents the Minkowski metric, $H$ is a scalar
function on spacetime, which decays as $~1/r$ at large radii, and
$l_\mu$ is null with respect to both the full metric and the Minkowski
metric (the concrete expressions can be found, e.g., in
\cite{Matzner-Huq-Shoemaker:1999}).  The Kerr-Schild form is preserved
under a Lorentz-transformation, and the metric for a boosted black
hole can be obtained by a suitable Lorentz-transformation on $l_\mu$.
In order to construct initial data for a spacetime containing two
black holes at coordinate locations $\vec c_{A,B}$ with masses
$M_{A,B}$, velocities $\vec v_{A,B}$ and spins $M_A\vec a_A$ and
$M_B\vec a_B$, one constructs the Kerr-Schild form Eq.~(\ref{eq:KS})
for each of the two black holes separately.  The free data for the
extrinsic curvature decomposition is then taken as the superposition
\begin{align}\label{eq:BKS-metric}
\cg_{ij}&=\delta_{ij}+2H^A l_i^A l_j^A + 2H^B l_i^B l_j^B,\\
\trK&=K_A+K_B,\\
\label{eq:BKS-exCurv}
\tilde M^{ij}&=\left(K^{(i}_{A\,k}+K^{(i}_{B\,k}
-\frac{1}{3}\delta^{(i}_{\,k}\left(K_A+K_B\right)\right)\cg^{j)k}.
\end{align}
Equation~(\ref{eq:BKS-metric}) is a natural generalization to two
black holes of Eq.~(\ref{eq:KS}).  Equation~(\ref{eq:BKS-exCurv}) is
somewhat complicated because $\tilde M^{ij}$ must be tracefree with
respect to the superposed metric $\cg_{ij}$,
cf. Eq.~(\ref{eq:Mij-decomposition}).  In
Ref.~\cite{Marronetti-Matzner:2000} the elliptic equations of the
extrinsic curvature decomposition were solved for the case
$\tilde\sigma\equiv 1$ (i.e. one of the old variants of the extrinsic
curvature decomposition).

In the limit of large separation of the black holes, one obtains two
widely separated Kerr-Schild metrics, each of which, by construction
will represent a (possibly) boosted and spinning black hole exactly,
so that the superposed Kerr-Schild approach certainly is advantageous
for widely separated binaries.  However, the most interesting initial
data is for a binary with separation close to the innermost stable
circular orbit, with black hole separations of only a few
Schwarzschild radii.  In that case, close to the horizon of, say, hole
$A$, the correction $2H^Bl^B_il^B_j$ due to hole $B$ will {\em not} be
small; for example, at the intersection of the line connecting the
centers $\vec c_A$ and $\vec c_B$ with the horizon of hole $A$, one
finds for separation $10M$ and for non-spinning black holes,
\begin{equation}\label{eq:HB-over-HA}
\frac{H^B}{H^A}=\frac{1}{4}.
\end{equation}
Given the nonlinearities in Einsteins equations, it is not clear how
the final initial data set will be influenced by this large
perturbation.  Since it is expected that only a few percent of the
energy will be emitted in gravitational waves during the inspiral
and merger of a binary black hole, it is necessary to control the
energy-content of the initial data set to at least one percent of the
energy (and preferably much better).  In view of the
ratio~(\ref{eq:HB-over-HA}), it seems unlikely that this is the case.
Indeed, the author~\cite{Pfeiffer-Cook-Teukolsky:2002} has examined
the proposal of superposed Kerr-Schild quantities extensively with a
spectral code for the construction of two black holes at rest.  Using
superposed Kerr-Schild quantities within the conformal thin sandwich
formalism and within the extrinsic curvature formalism resulted in
variations of the ADM-energy by up to several per cent.  It was also
found that the resulting initial data sets depend sensitively on the
choice of extrinsic curvature: Removal of the longitudinal part of
$\tilde M^{ij}$ by a procedure analogous to
Eqs.~(\ref{eq:Mij-decomposition})--(\ref{eq:Aij-from-Mij}) before
solving the constraint equations changes the ADM-energy by
several percent.

We also note that superposed Kerr-Schild metrics singles out
arbitrarily a specific slicing: {\em Any} slicing of the Kerr
spacetime will give rise to three-metric and extrinsic curvatures
which can be superposed similarly to
Eqs.~(\ref{eq:BKS-metric})--(\ref{eq:BKS-exCurv}).  Furthermore, the
superposition depends on the {\em spatial} coordinate system used to
represent the single black hole quantities.  For non-spinning,
unboosted black holes, the spatial part of the Kerr-Schild metric is
spherically symmetric, and can be made conformally flat by a radial
coordinate transformation~\cite{Cook:2002}.  After this spatial
coordinate transformation, the equivalent of Eq.~(\ref{eq:BKS-metric})
is superposition of two conformally flat metrics, which implies that
$\cg_{ij}$ should be chosen to be conformally
flat~\cite{Pfeiffer:2003, Cook-Pfeiffer:2004}.  While superposing
single black hole quantities is certainly an interesting route, more
exhaustive investigations into their properties are necessary.

Conformal flatness is also questionable because a {\em binary} compact
object is not conformally flat at second post-Newtonian
order~\cite{Rieth:1997}.  Tichy {\em et
al.}~\cite{Tichy-Bruegmann-etal:2003} address this issue by using
post-Newtonian results to set the free data for the extrinsic
curvature decomposition: The post-Newtonian expansion of the spatial
metric can be written as~\cite{Jaranowski-Schaefer:1998}
\begin{equation}\label{eq:PN}
\g_{ij}^{\rm PN}=\CF^4_{\rm PN}\delta_{ij}+h_{ij}^{TT},
\end{equation}
where $\CF_{\rm PN}$ and $h_{ij}^{TT}$ are given as expansions in the
velocity $v/c$.  The spatial metric Eq.~(\ref{eq:PN}) and the
corresponding extrinsic curvature are known for general motion of the
masses, i.e. in particular for a binary in circular orbits.  The
extrinsic curvature can be similarly expressed as a series in $v/c$,
the leading order term being the Bowen-York momentum.  One now bases
the conformal metric on Eq.~(\ref{eq:PN}), and similarly for the
extrinsic curvature.  This approach is the only one so far, which, in
principle, can account for genuine binary contributions to the free
data.

Tichy {\em et al.}~\cite{Tichy-Bruegmann-etal:2003} encountered two
important issues during implementation of this approach.  The first
problem is that close to the point-masses, the post-Newtonian
expansion breaks down, that is, different terms in the expansion grow
with different inverse powers of distance to the point mass.  Terms of
higher order in $v/c$ diverge faster than the lower order terms, so
that the choice which terms one retains (e.g., consistent in the order
$v/c$) influences the obtained initial data sets strongly.  This may
reflect a fundamental problem of using post-Newtonian expansions: They
are least accurate close to the point-masses.  The second problem
encountered by Ref.~\cite{Tichy-Bruegmann-etal:2003} is related to the
fact that $\CF_{\rm PN}$ has a singularity at the point-masses.  In
order to obtain a finite conformal metric, the authors decided to
conformally scale $\g_{ij}^{\rm PN}$,
\begin{equation}
\cg_{ij}\equiv \Omega^{-4}g_{ij}^{\rm PN},
\end{equation}
where the leading order behavior of $\Omega$ close to the
singularities is identical to that of $\CF_{\rm PN}$ However, the
concrete choice for $\Omega$ influences the resulting initial data
sets in a significant way; for example, the simplest choice
$\Omega=\CF_{\rm PN}$ leads to binary black hole initial data sets
with {\em increasing} energy as the separation between the black holes
is reduced.  Tichy {\em et al.} do not employ the extrinsic curvature
decomposition presented in Sec.~\ref{sec:IVP:ExCurv}, but rather the
old version without the weight-function $\sigma$ (their work predates
discovery of the weight-function in Ref.~\cite{Pfeiffer-York:2003}).
This old decomposition is not invariant to conformal transformations
of the free data, while the new one is
[cf. Eq.~(\ref{eq:freedata-scaled2})], so that ambiguities related to
the choice of $\Omega$ arise only in the old decomposition.  It would
be interesting to repeat Tichy's work with the new decomposition.

\subsection{Numerical solutions of the conformal thin sandwich equations}

The conformal thin sandwich formalism has two advantages over the
extrinsic curvature decomposition.  First, it replaces the free data
corresponding to the tracefree extrinsic curvature $\tilde M^{ij}$ and
the weight function $\tilde\sigma$ by freely specifiable
time-derivatives, $\partial_t\cg_{ij}\!=\!\cu_{ij}$ and
$\partial_t\trK$.  Time-derivatives allow for easier physical
interpretation of the initial data set under construction, and as we
will see below, in the most interesting case, there is a natural
choice for these time-derivatives, thus avoiding ambiguities related
to the choice of, e.g., $\tilde M^{ij}$ in Eq.~(\ref{eq:BKS-exCurv}).
As second advantage, solution of the conformal thin sandwich equations
results directly in a preferred gauge choice $(\N, \beta^i)$, which
can be used in evolutions of the initial data set (at least during the
initial stages of the evolution).

The orbits of binary compact objects of similar masses are expected to
circularize before late inspiral, so that shortly before merger, the
compact objects will move in circular orbits about each other.
Therefore, initial data with two black holes in such a quasi-circular
orbit is of particular interest.  In the co-rotating coordinate system
such a configuration will appear time-independent (up to corrections
due to radiation reaction which are neglected so far), and so the free
data corresponding to time-derivatives is simply set to zero.  This
motivates use of the conformal thin sandwich formalism with the free
data Eq.~(\ref{eq:CTS-freedata2}), fully half of which consists of
time-derivatives:
\begin{equation}
\label{eq:BBH-freedata}
\cu_{ij}=\partial_t\trK=0,
\end{equation}

The notion of time-independence in the corotating frame is basically
equivalent to the assumption of an approximate helical Killing vector
\cite{Gourgoulhon-Grandclement-Bonazzola:2001a}, which close to
infinity takes the form $l_{\rm HKV}=\partial/\partial t_0 + \Omega_0
\partial/\partial\phi_0$, where $\partial/\partial t_0$ and
$\partial/\partial\phi_0$ are, respectively, the asymptotic timelike
and rotational Killing vectors.  This follows simply from the fact
that, in the corotating frame, the time-vector $t^\mu$ of the
evolution coincides with the approximate helical Killing vector.
Assuming asymptotic flatness, and requiring that the time-vector of a
subsequent evolution coincides with $l_{\rm HKV}$ close to infinity
leads in boundary conditions at the outer boundary,
\begin{align}
\CF&\to 1, &r\to \infty,
\\ \beta^i&\to \Omega_0\left(\frac{\partial}{\partial\phi_0}\right)^i,
&r\to\infty,\\
N&\to 1,&r\to\infty.
\end{align}

It now remains to choose $\cg_{ij}$ and $\trK$ as well as inner
boundary conditions at excision spheres around the singularities of
the black holes\footnote{The puncture
method~\cite{Brandt-Bruegmann:1997} avoids inner boundaries in the
computational domain, but it cannot be generalized to the construction
of quasi-equilibrium initial data within the conformal thin sandwich
formalism \cite{Hannam-Evans-etal:2003}.}.  

The first proposal for such boundary conditions was made by
Gourgoulhon {\em
et al.}~\cite{Gourgoulhon-Grandclement-Bonazzola:2001a}\footnote{Gourgoulhon
{\em et al.} scale the tracefree extrinsic curvature differently from
Eq.~(\ref{eq:Aij-scaling}), so that the elliptic equations differ in
lower order terms and some of the statements made in
Sec.~\ref{sec:IVP:CTS} do not apply.}, and numerical solutions were
obtained with a spectral code in a companion
paper~\cite{Grandclement-Gourgoulhon-Bonazzola:2001b}.
Refs.~\cite{Gourgoulhon-Grandclement-Bonazzola:2001a,Grandclement-Gourgoulhon-Bonazzola:2001b}
assumed conformal flatness and maximal slicing to simplify the
problem.  Part of the necessary boundary conditions at inner
boundaries were derived from the demand that the obtained initial data
be inversion symmetric across the throats of the horizons, resulting
in a Robin boundary condition on the conformal factor,
\begin{equation}\label{eq:GGB-PsiBC}
\tilde s^i\tilde\nabla_i\CF+\frac{\CF}{2r}=0,\qquad\mbox{on $\cal S$}
\end{equation}
where $r$ denotes the coordinate radius of each excised sphere, and
$\tilde s^i$ represents the outward pointing (i.e. toward infinity)
normal to ${\cal S}$, tangent to the hypersurface and normalized such
that $\cg_{ij}\tilde s^i\tilde s^j=1$.  Inversion symmetry also
implies that the lapse function must vanish on the throat,
\begin{equation}\label{eq:GGB-LapseBC}
N=0,\qquad\mbox{on $\cal S$},
\end{equation}
which constitutes a boundary condition on the fifth initial value
equation, Eq.~(\ref{eq:dtK4}).  Boundary conditions on the shift,
finally, were obtained by requiring that the time-vector $t^\mu=N
n^\mu+\beta^\mu$ be tangent to the generators of the horizon.  In
particular, $t^\mu$ must be null, $t^\mu t^\nu \gfour_{\mu\nu}=0$, so
that from Eq.~(\ref{eq:GGB-LapseBC}) it follows
\begin{equation}
\beta^i=0,\qquad\mbox{on $\cal S$}.
\end{equation}
Since the lapse-function vanishes on ${\cal S}$, inspection of
Eqs.~(\ref{eq:CTS-Aij}) and~(\ref{eq:BBH-freedata}) reveals that the
extrinsic curvature will be finite on $\cal S$ only if
$(\Long\beta)^{ij}$ vanishes.  However, this is not the case, so that
Refs.~\cite{Gourgoulhon-Grandclement-Bonazzola:2001a,
Grandclement-Gourgoulhon-Bonazzola:2001b} must resort to a
regularization procedure which causes the resulting initial data 
to {\em violate} the constraints~\cite{Cook:2002} on some small
level.  The authors continue and construct a sequence of
quasi-circular orbits for corotating black holes, and find that the
location of the innermost stable circular orbit (ISCO) is close to
post-Newtonian predictions.

A more general and more sophisticated approach to inner boundary
conditions is due to Cook \& Pfeiffer \cite{Cook:2002, Pfeiffer:2003,
Cook-Pfeiffer:2004}.  The idea is to use the {\em physical} concept of
black holes in quasi-equilibrium to derive as many boundary conditions
as possible.  As we will see, this approach can be used in combination
with any choice for $\cg_{ij}$ and $\trK$.  Two topological spherical
regions with boundary $\cal S$ are excised from the computational
domain. Denote the outward-pointing null-geodesics tangent to ${\cal
S}$ by $k^\mu$, and their expansion by $\theta$.  Then the following
demands are made:
\begin{enumerate}
\item The surfaces $\cal S$ must be apparent horizons\footnote{More
precisely, marginally outer-trapped surfaces.},
i.e. $\left.\theta\right|_{\cal S}=0$.
\item The shear of $k^\mu$ must vanish on ${\cal S}$.
\item When the initial data are evolved with the lapse and shift
obtained during the solution of the conformal thin sandwich equations,
then the {\em coordinate surfaces} of the apparent horizons are to
remain stationary initially.
\end{enumerate}
The first of these demands simply localizes the apparent horizon in
coordinate space, so that it is known where to apply the other
conditions.  The second requirement is based on quasi-equilibrium;
it implies, using Raychaudhuri's equation for null-congruences, that
\begin{equation}
\left.{\cal L}_k\theta\right|_{\cal S}=0.
\end{equation}
That is, initially, the apparent horizon will evolve along $k^\mu$ and
because the expansion of $k^\mu$ vanishes, the apparent horizon area
will remain constant initially.  The third demand is a gauge choice
ensuring that the coordinates are adapted to the physical situation.
Reexpressing demands (1) and (3) in the variables of the conformal
thin sandwich equations yields the following conditions:
\begin{align}
\label{eq:AH_BC}
\tilde s^k\cderiv_k\ln\CF &= -\frac14\tilde h^{ij}\cderiv_i\tilde
s_j +\frac{1}{6}\CF^2 \trK-\frac{\CF^2}{8\N}\tilde s_i\tilde
s_j(\cLong\beta)^{ij},
&&\mbox{on $\cal S$,}\\
\label{eq:Shift_BC}
\beta^i &= \psi^2\N\tilde  s^i+\beta^i_{\parallel},
&&\mbox{on $\cal S$,}
\end{align}
where $\tilde s^k$ is defined below Eq.~(\ref{eq:GGB-PsiBC}), $\tilde
h_{ij}=\cg_{ij}-\tilde s_i\tilde s_j$ denotes the induced conformal
metric on $\cal S$ and $\beta^i_{\parallel}$ is the projection of
$\beta^i$ into $\cal S$ (i.e. $\beta^i_{\parallel}\tilde s_i=0$).
Equation~(\ref{eq:AH_BC}) represents a nonlinear Robin-like boundary
condition on $\CF$, whereas Eq.~(\ref{eq:Shift_BC}) is essentially a
Dirichlet condition on the shift.  The remaining demand of vanishing
shear of $k^\mu$, finally, implies~\cite{Cook-Pfeiffer:2004} that the
tangential component of the shift $\beta^i_{\parallel}$ must be a
conformal Killing vector of the two-dimensional surface ${\cal S}$,
\begin{equation}\label{eq:Shift_perp}
\tilde D^{(i}\beta_{\parallel}^{j)}
- \mbox{$\frac12$}\tilde h^{ij}\tilde D_k\beta_{\parallel}^k = 0
\qquad\qquad\mbox{on $\cal S$,}\end{equation}
where $\tilde D_i$ denotes the covariant derivative within ${\cal S}$,
which is induced by the conformal metric.  Cook \& Pfeiffer point out
that solutions to Eq.~(\ref{eq:Shift_perp}) can be found {\em before}
the conformal thin sandwich equations are solved: Each connected
component of $({\cal S}, \tilde h_{ij})$ is topologically $S^2$ and
therefore conformally equivalent to a unit 2-sphere embedded in
three-dimensional Euclidean space.  For such a Euclidean 2-sphere,
Euclidean rotations around the center represent Killing vectors, which
are conformal Killing vectors of any conformally related manifold,
including $({\cal S}, \tilde h_{ij})$.  Therefore, these rotations
will satisfy Eq.~(\ref{eq:Shift_perp}).  The freedom to specify
arbitrary rotations can be used to construct black holes with
arbitrary rotational state.  Interestingly, the lapse boundary
condition is {\em not} determined by quasi-equilibrium considerations,
but is rather part of the temporal gauge choice.

The quasi-equilibrium boundary conditions at the inner excision
regions allow for arbitrary specification of conformal metric, mean
curvature, and shape of the excision regions.  Spectral numerical
solutions of the conformal thin sandwich equations (with lapse
equation) using the quasi-equilibrium boundary conditions
Eqs.~(\ref{eq:AH_BC})--(\ref{eq:Shift_perp}) were obtained in
\cite{Cook-Pfeiffer:2004} for the special case of a conformally flat
metric and excision of exact coordinate spheres.  Three lapse boundary
conditions at the excised regions (Dirichlet, von Neumann and Robin),
combined with irrotational and corotating black holes, and combined
with two choices for the mean curvature $\trK$ were explored, for a
total of 12 sequences of quasi-circular orbits.  Furthermore, spinning
and boosted single black holes were constructed under the
approximation of conformal flatness, and it was shown that the errors
introduced by the use of a flat conformal metric are fairly small.
Nonetheless, the freedom in the choice of $\cg_{ij}, \trK, {\cal S}$
and the lapse boundary condition can be used to further fine tune the
method.

There is a close connection between the quasi-equilibrium inner
boundary conditions and the isolated horizon framework developed by
Ashtekar and coworkers (see,
e.g. \cite{Ashtekar-Beetle-etal:2000,Dreyer-Krishnan-etal:2003,
Ashtekar-Krishnan:2003}).  Jaramillo~{\em
et al.}~\cite{Jaramillo-Gourgoulhon-Marugan:grqc0407063} work out
this relation very clearly.  Not surprisingly their results are closely
related to findings in Ref.~\cite{Cook-Pfeiffer:2004}, including an
independent argument why the lapse boundary condition is not
determined by quasi-equilibrium/isolated horizon considerations.
Ref.~\cite{Jaramillo-Gourgoulhon-Marugan:grqc0407063} does not present
numerical results, and it is not immediately clear how to translate
the conditions in this paper into a form usable in numerical
simulations.  Yo {\em el. al.} \cite{Yo-Cook-etal:grqc0406020}
recently solved the five coupled conformal thin sandwich equations on
a Kerr-Schild background, however, with much simpler inner boundary
conditions, and with much less accuracy, owing to the employed
finite-difference code with uniform grid spacing.

As a final, somewhat unrelated application of the conformal thin
sandwich equations, we mention construction of initial data with
superposed gravitational waves~\cite{Pfeiffer-Kidder-etal:grqc0410016}
(see
also~\cite{Shibata-Nakamura:1995,Bonazzola-Gourgoulhon-etal:2004}).
The idea is very simple: Given a slice through a stationary background
spacetime with induced metric $g^0_{ij}$ and mean curvature $\trK^0$,
as well as a linearized gravitational wave $h_{ij}$ (either on the
background, or on Minkowski space), set the free data for the
conformal thin sandwich equations as
\begin{align}
\cg_{ij}&=g^0_{ij} + A h_{ij},\\
\cu_{ij}&= A \partial_t h_{ij},\\
\trK&=\trK^0,\\
\partial_t\trK&=0,
\end{align}
where $A$ is the adjustable amplitude of the perturbation.  In
Ref.~\cite{Pfeiffer-Kidder-etal:grqc0410016} the conformal thin
sandwich equations were solved with these free data (and Dirichlet
boundary conditions).  Solutions were obtained for perturbations of
Minkowski space and of a Schwarzschild black hole with very large
amplitudes.  These perturbed black hole initial data are used to test
constraint preserving boundary conditions for evolutions of Einstein's
equations~\cite{Kidder-Lindblom-etal:2004}.

\section{Conclusion}

We summarized the conformal method to solve the constraints of general
relativity, both in its Lagrangian and Hamiltonian viewpoints.  By
virtue of the recently discovered {\em weighted} transverse-traceless
decomposition in the Hamiltonian viewpoint (the extrinsic curvature
formulation), it is shown to be completely equivalent to the
Lagrangian viewpoint (the conformal thin sandwich formulation).  Both
pictures are invariant to conformal transformations of the free data,
and in both pictures, the decomposition of the tracefree extrinsic
curvature commutes with the conformal scalings.

Subsequently, we summarized recent numerical work on solving the
coupled constraint equations, in either viewpoint, concentrating on
quasi-equilibrium solutions using the conformal thin sandwich
equations, as well as superposed Kerr-Schild data and initial data
based on post-Newtonian results using the extrinsic curvature
formulation.

\section*{Acknowledgments}

I am grateful for helpful discussions with James York, Lee Lindblom and
Gregory Cook.  This work was supported by a Sherman Fairchild Prize
fellowship and by NSF grants PHY-0244906 and PHY-0099568 to the
California Institute of Technology.

\end{document}